\documentclass[twocolumn,showpacs,preprintnumbers,amsmath,amssymb,superscriptaddress,prl]{revtex4}
\usepackage{epsfig}
\usepackage{graphicx}
\usepackage{color}
\usepackage{bm}

\begin{document}

\title{Relativistic N\'eel-order fields induced by electrical current in antiferromagnets}

\author{J.~\v{Z}elezn\'y$^\ast$}
\affiliation{Institute of Physics ASCR, v.v.i., Cukrovarnick\'a 10, 162 53
Praha 6, Czech Republic}
\affiliation{Faculty of Mathematics and Physics, Charles University in Prague,
Ke Karlovu 3, 121 16 Prague 2, Czech Republic}
\author{H.~Gao$^\ast$}
\affiliation{Department of Physics, Texas A\&M University, College Station, Texas 77843-4242, USA}
\author{K. V\'yborn\'y}
\affiliation{Institute of Physics ASCR, v.v.i., Cukrovarnick\'a 10, 162 53
Praha 6, Czech Republic}
\author{J.~Zemen}
\affiliation{Department of Physics, Blackett Laboratory, Imperial College London, London SW7 2AZ, United Kingdom}
\author{J.~Ma\v{s}ek}
\affiliation{Institute of Physics ASCR, v.v.i., Na Slovance 2, 182 21 Praha 8, Czech Republic}
\author{A.~Manchon}
\affiliation{Physical Science and Engineering Division, King Abdullah University of Science and Technology (KAUST), Thuwal 23955-6900, Kingdom of Saudi Arabia}

\author{J.~Wunderlich}
\affiliation{Institute of Physics ASCR, v.v.i., Cukrovarnick\'a 10, 162 53 Praha 6, Czech Republic}
\affiliation{Hitachi Cambridge Laboratory, Cambridge CB3 0HE, United Kingdom}
\author{J.~Sinova}
\affiliation{Institut f\"ur Physik, Johannes Gutenberg Universit\"at Mainz, 55128 Mainz, Germany}
\affiliation{Institute of Physics ASCR, v.v.i., Cukrovarnick\'a 10, 162 53
Praha 6, Czech Republic}
\affiliation{Department of Physics, Texas A\&M University, College Station, Texas 77843-4242, USA}
\author{T.~Jungwirth}
\affiliation{Institute of Physics ASCR, v.v.i., Cukrovarnick\'a 10, 162 53
Praha 6, Czech Republic} 
\affiliation{School of Physics and
Astronomy, University of Nottingham, Nottingham NG7 2RD, United Kingdom}

\begin{abstract}
We predict  that a lateral electrical current  in antiferromagnets can induce non-equilibrium N\'eel-order fields, i.e. fields whose sign alternates between the spin sublattices, which can trigger ultra-fast spin-axis reorientation. Based on microscopic transport theory  calculations we identify staggered current-induced fields analogous to the intra-band and to the intrinsic inter-band spin-orbit fields previously reported in ferromagnets with a broken inversion-symmetry crystal. To illustrate their rich physics and utility, we considered bulk Mn$_2$Au  with the two spin sublattices forming inversion partners, and  a 2D square-lattice antiferromagnet with broken structural inversion symmetry modeled  by a Rashba spin-orbit coupling. We propose an AFM memory device with electrical writing and reading. 
\end{abstract}


\maketitle

Commercial spin-based memory  and storage 
devices rely on one type of magnetic order, ferromagnetism, and one basic principle, that the opposite 
spin orientations in a ferromagnet (FM) represent the 0's and 1's \cite{Chappert:2007_a}.  Magnetic random access memory (MRAM) is a solid-state-memory variant of the hard disk where the magnetic medium for storing and the magnetoresistive read-element are merged into one. 
Unlike in hard disks, the magnetic stray field of the FM is not used for reading  
the FM bit and the latest 
spin--torque based MRAMs  do not even use magnetic fields coupled to the FM moment 
for writing 
\cite{note1}. From this it appears natural  to consider  antiferromagnets (AFMs) as active building blocks of spintronic devices, where magnetic order is accompanied by a zero net magnetic moment 
\cite{Nunez:2006_a,Shick:2010_a,Jungwirth:2010_a}. 

Antiferromagnets are attractive for spintronics because they offer insensitivity to magnetic field perturbations, 
produce no perturbing stray fields,  are readily compatible with metal, semiconductor, or insulator electronic structure, can act as a magnetic memory, and
can generate large magneto-transport effects \cite{Shvets:2006_a,Park:2010_a,Marti:2013_a}.
For example, two distinct stable states of an AFM with orthogonal AFM spin-axis directions were set in an FeRh ohmic resistor and shown to be insensitive to fields as high as 9 T at ambient conditions \cite{Marti:2014_a}.   A $\sim1$\% AFM anisotropic magnetoresistance (AMR) was used to electrically detect the states, in complete analogy to the $\sim1$\% AMR of NiFeCo based bits in the first generation of FM-MRAMs \cite{Daughton:1992_a}. 
Also in analogy with the development of 
FM spintronics, very large ($\sim100$\%) magnetoresistance  signals were reported in AFM tunnel devices~\cite{Park:2010_a}. 

The AFM N\'eel-order spin-axis direction can be controlled indirectly by a magnetic field via an attached exchange-coupled FM \cite{Park:2010_a,Marti:2013_a} or, without the auxiliary FM, 
by techniques analogous to heat-assisted magnetic recording \cite{Petti:2013_a,Marti:2014_a}. 
However, as with  heat-assisted FM-MRAM  \cite{note2,Dieny:2010_a}, the speed and energy 
efficiency of this  method are limited.  
Here we predict a novel mechanism for AFM spin-axis reorientation by a lateral electrical current via 
N\'eel-order spin-orbit torque (NSOT) fields, {\it i.e.}, via non-equilibrium fields that alternate in sign between the two spin sublattices. This relativistic mechanism  does not involve  FMs, heating, or magnetic fields, and offers ultra-short times unparalleled in FMs. 

The microscopic origin of our NSOT fields is analogous to the relativistic spin-orbit torques (SOTs) observed recently in magnets with broken bulk or structural inversion symmetry \cite{Bernevig:2005_c,Manchon:2008_b,Manchon:2009_a,Matos:2009_a,Chernyshov:2009_a,Garate:2009_a,Endo:2010_a,Fang:2010_a,Miron:2010_a,Miron:2011_b,Liu:2012_a,Li:2013_a,Garello:2013_a,Kurebayashi:2013_a},
and is distinct in origin from the non-relativistic spin-transfer torques \cite{Ralph:2007_a,Nunez:2006_a,Gomonay:2014_a}.
We demonstrate below   two types of NSOTs in two model systems. 

\begin{figure}[h]
\hspace*{0cm}\epsfig{width=.9\columnwidth,angle=0,file=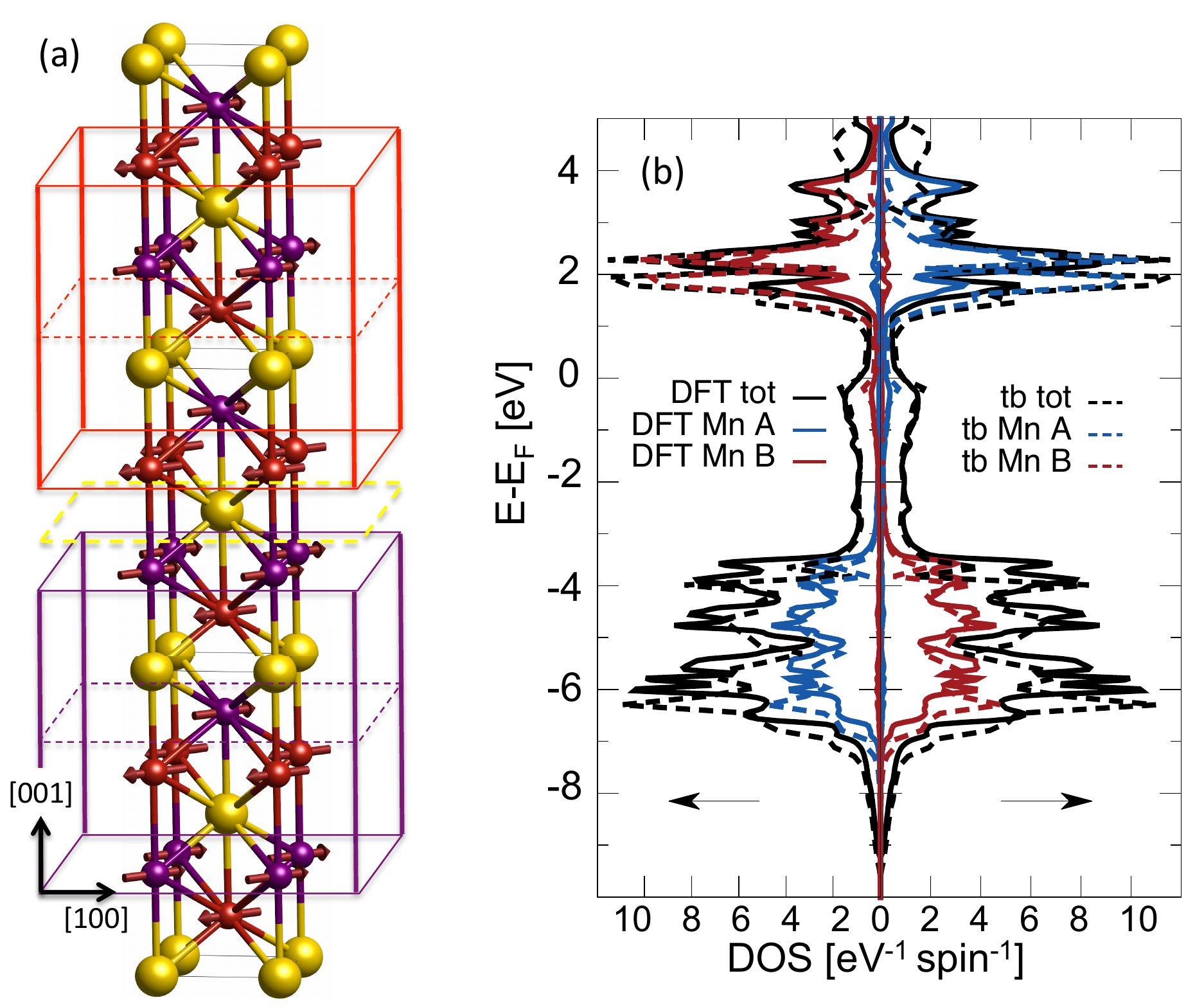}
%
\caption{(a) Mn$_2$Au crystal structure and antiferromagnetic ordering. The two spin-sublattices have broken inversion symmetry as illustrated by the red and purple colors. The full crystal is centrosymmetric around the Au atom as also highlighted in the figure. (b) Total, sublattice, and spin projected density of states from the {\em ab-initio} calculation and for the
tight-binding Hamiltonian model.}
\label{Mn2Au}
\end{figure}

A field-like NSOT appears in Mn$_2$Au AFM  \cite{Shick:2010_a,Wu:2012_a,Barthem:2013_a}, whose MoSi$_2$-type bct structure and AFM 
ordering are shown  in Fig.~\ref{Mn2Au}(a). It is analogous to the field-like SOT arising from the inverse spin galvanic effect \cite{Bernevig:2005_c,Manchon:2008_b,Manchon:2009_a,Matos:2009_a,Chernyshov:2009_a,Garate:2009_a,Endo:2010_a,Fang:2010_a,Miron:2010_a,Li:2013_a}, observed previously in broken inversion-symmetry 
paramagnets or FMs. However, Mn$_2$Au is bulk centrosymmetric and the current-induced NSOT arises from the fact that the lattice can be divided into two sublattices, which, individually, have broken inversion symmetry and form inversion partners  \cite{Zhang:2014_a}. Each sublattice gives opposite inverse spin galvanic effects, resulting in the NSOT field.  The range of 
materials in which the relativistic current-induced torques can occur is therefore not restricted to FMs and, moreover, is not restricted to crystals  with global broken inversion symmetry. In Mn$_2$Au, the inversion partner sublattices coincide with the two AFM spin sublattices which makes the material an attractive candidate for observing the NSOT. 

In AFMs where the two spin-sublattices do not form inversion partners  a NSOT can still occur. We illustrate it below
in a 2D square lattice where the same broken inversion symmetry term in the Hamiltonian is shared by both spin sublattices.  
Here the resulting NSOT is analogous to the intrinsic anti-damping SOT recently observed in broken bulk inversion symmetry FMs
\cite{Kurebayashi:2013_a}.

{\it Models and methods:}
In Mn$_2$Au we diagonalized a microscopic multi-orbital tight-binding Hamiltonian to obtain the energy spectrum and eigenfunctions used in our transport calculations. The form of the tight-binding Hamiltonian was obtained following the procedure for bimetallic alloys described in Ref.~\cite{Zemen:2014_a}. The accuracy of the tight-binding energy spectrum is confirmed in Fig.~\ref{Mn2Au}(b) by comparing the electronic structure to the {\em ab initio} density-functional theory (DFT) calculations. 

\begin{figure}[h]
\hspace*{0cm}\epsfig{width=0.9\columnwidth,angle=0,file=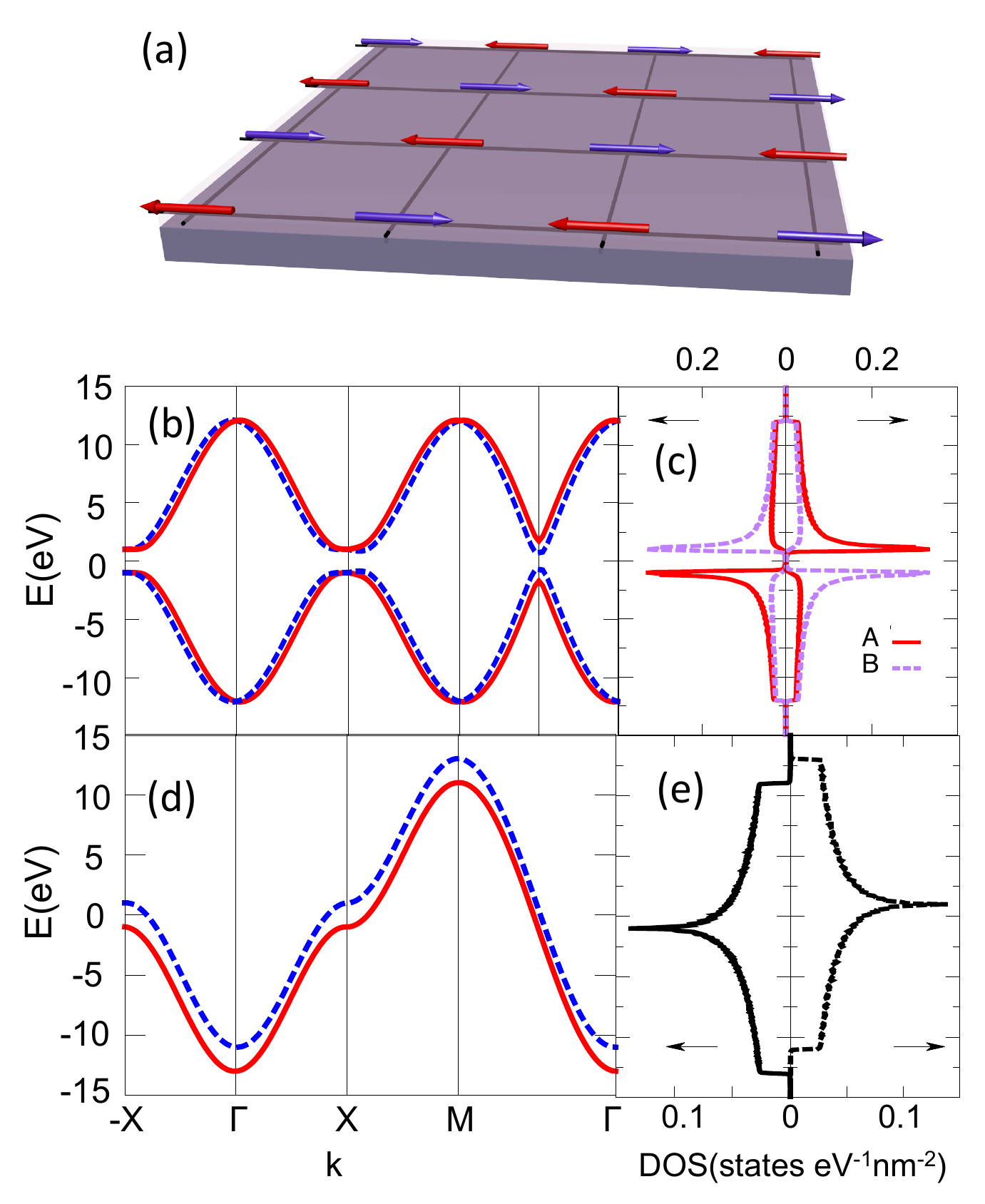}
%
\caption{(a) 2D AFM square lattice model with Rashba spin-orbit coupling. (b) and (c) Band structure and the spin-resolved density of states projected in each sublattice for the AFM state. (d) and (e) Band structure and the spin-resolved density of states for the FM state. Here the hopping parameter $t_N=3.0$ eV, $J_{sd}=1.0$ eV, and $V_{so}=0.1$ eV.}
\label{2DRashba}
\end{figure}

The other model structure comprises a 2D AFM square lattice with Rashba spin-orbit coupling  due to the broken structural inversion symmetry and is relevant, e.g., to common experimental geometries in which a thin AFM film is interfaced with another layer. The model is sketched in Fig.~\ref{2DRashba}(a) and its Hamiltonian is given by
\begin{equation}
 H = \sum_{<ij>}J_{dd} \vec{S_i}\cdot\vec{S_j} + H^{tb} + H_R+ \sum_i J_{sd}\vec{s}\cdot\vec{S_i} \,.
\end{equation}
Here $J_{dd}$ is the local moment 
exchange constant, $J_{sd}$ is the local moment -- carrier  exchange constant, $H^{tb}$ is the tight binding Hamiltonian for the carriers,  and $H_R$ is the Rashba spin-orbit interaction in a 2D system,  given by
\begin{eqnarray}
H_R &=& V_{SO}\sum_i [(c_{i\uparrow}^{\dagger} c_{i+\delta_x\downarrow} - c_{i\downarrow}^{\dagger} c_{i+\delta_x\uparrow}) 
\nonumber\\&&- i(c_{i\uparrow}^{\dagger} c_{i+\delta_y\downarrow} + c_{i\downarrow}^{\dagger} c_{i+\delta_y\uparrow}) + \mbox{H.c.}],
\end{eqnarray}
where $V_{SO}$ represents the spin-orbit coupling strength, and $\delta_x$, $\delta_y$ label the nearest neighbors 
direction. 

The current-induced nonequilibrium spin density $\delta\vec{s}$ can be calculated via the Kubo linear response \cite{Garate:2009_a}:
\begin{equation}
\delta\vec{s}= \frac{\hbar }{2\pi L^2 }\mbox{Re}\sum_{\vec{k}\alpha\beta}(\vec{s})_{\alpha\beta}(e\vec{E}\cdot \vec{v})_{\beta\alpha}[G_{\vec{k}\alpha}^AG_{\vec{k}\beta}^R-G_{\vec{k}\alpha}^RG_{\vec{k}\beta}^R],
\end{equation}
where the Green's functions are $G_{\vec{k}\alpha}^R(E)|_{E=E_F}\equiv G_{\vec{k}\alpha}^R = 1/(E_F-E_{\vec{k}\alpha} + i\Gamma)$, with the property $G^A = (G^R)^*$. Here, $L$ is the dimension of the 2D system, $e$ is the charge of electron, $\vec{E}$ is the applied electric field, $E_F$ is the Fermi energy, $E_{\vec{k}\alpha}$ is the energy spectrum, and $\Gamma$ is the spectral broadening that models the effect of disorder. For small $\Gamma$, we can separate the total $\delta\vec{s}$
into the intra-band and inter-band contributions, with the intra-band term given
by
\begin{equation}\label{Boltzmann}
  \delta \vec{s}^{intra}=\frac{eE\hbar}{2\Gamma}\int\frac{d^3k}{(2\pi)^3}
  \sum_\alpha (\vec{s})_{\vec{k}\alpha}
   (v_I)_{\vec{k}\alpha}
  \delta(E_{\vec{k}\alpha}-E_F).
\end{equation}
Here  $(\vec{s})_{\vec{k}\alpha}$ denotes the
expectation value of the carrier spin, and $ (v_I)_{\vec{k}\alpha}$ the velocity component along the current
direction. This intra-band contribution in the Kubo formalism is equivalent to the Boltzmann transport theory expression  \cite{Bernevig:2005_c,Manchon:2008_b,Manchon:2009_a,Garate:2009_a,Fang:2010_a} and, similar to the charge conductivity, $\delta \vec{s}^{intra}\sim1/\Gamma$.

The inter-band contribution  dominating in the clean limit of $\Gamma\rightarrow 0$ is given by  \cite{Garate:2009_a},
\begin{eqnarray}
\delta \vec{s}^{inter} &=& \frac{\hbar}{L^2}\sum_{\vec{k}\alpha\neq\beta} (f_{\vec{k}\alpha}-f_{\vec{k}\beta})\mbox{Im}[(\vec{s})_{\alpha\beta}(e\vec{E}\cdot \vec{v})_{\beta\alpha}]
\nonumber\\&&\times
\frac{(E_{\vec{k}\alpha}-E_{\vec{k}\beta})^2-\Gamma^2}{[(E_{\vec{k}\alpha}-E_{\vec{k}\beta})^2+\Gamma^2]^2}\,.
\label{inter}
\end{eqnarray}
Here, the labels $\alpha$ and $\beta$ correspond to different bands, and $f_{\vec{k}\alpha,\beta}$ is Fermi distribution function.

\begin{figure}[h]
\hspace*{0cm}\epsfig{width=.9\columnwidth,angle=0,file=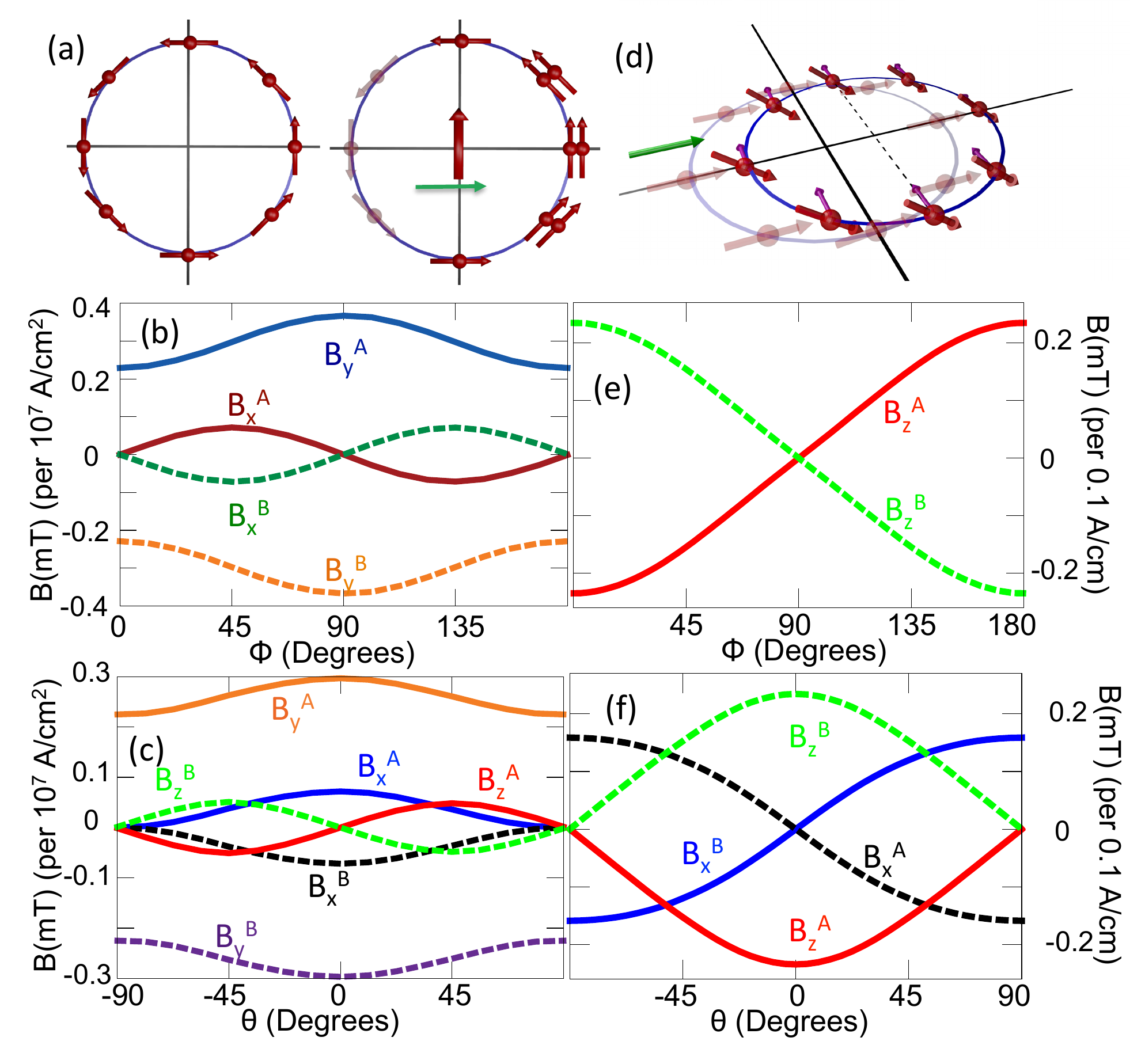}
%
\caption{(a) Schematics of the intra-band, inverse spin galvanic effect in a model Rashba system. Left panel represents equilibrium distribution of spins (red arrows), right panel shows the non-equilibrium redistribution resulting in a net in-plane spin-polarization (thick red arrow) perpendicular to current (green arrow). (b) Intra-band NSOT field in Mn$_2$Au as a function of the in-plane spin-axis angle. The sublattice index  A or B and component of the field x, y, or z ([100], [010], [001]) are shown for each curve.  (c) Same as (b) for the out-of-plane spin-axis angle. (d) Schematics of the intrinsic inter-band contribution to the non-equilibrium spin polarization. In equilibrium all spins are approximately aligned with the exchange field which is considered to be stronger than the Rashba field. A non-equilibrium in-plane Rashba field (purple arrows) aligned perpendicular to the applied current causes and out-of-plane tilt of the carrier spins on the shifted Fermi surface. (e),(f) Inter-band NSOT fields as a function of spin-axis angles in the 2D Rashba AFM for $\Gamma=0.01$~eV and $E_F=-2$~eV. Other parameters of the model are as in Fig.~\ref{2DRashba} In all panels current is along the [100]-axis.
}
\label{SOT_angle}
\end{figure}
{\it Results in Mn$_2$Au:} In Figs.~\ref{SOT_angle}(b),(c) we show the $\Gamma$-independent intra-band NSOT field per applied current for Mn$_2$Au. It is evaluated from Eq.~(\ref{Boltzmann}) and projected on each sublattice, assuming AFM spin-axis rotation in the [100]-[010] plane  ($\phi=0$ corresponds to the [100] spin-axis direction) and in the [110]-[001] plane ($\theta=0$ corresponds to the [110] easy-spin-axis in Mn$_2$Au). Current is applied along the [100]-direction and the NSOT field is obtained from 
the non-equilibrium spin density considering a typical exchange-coupling energy scale in transition metals $\sim 1$~eV \cite{Haney:2013_a}. 

NSOT fields  on each sublattice are non-zero and have opposite sign. 
The largest component  is in the $[100]-[010]$ plane in the direction perpendicular to the applied current for all AFM spin-axis directions.  The magnitude of the NSOT field in the Mn$_2$Au AFM is comparable to the counterpart SOT fields observed in FM transition metal structures. Note that for current along the $[001]$ direction the resulting NSOT field is zero.

The results imply that this intra-band NSOT is an AFM counterpart of the inverse spin galvanic effect \cite{Edelstein:1990_a}, or the intra-band, field-like, SOT \cite{Bernevig:2005_c,Manchon:2008_b,Manchon:2009_a,Matos:2009_a,Chernyshov:2009_a,Garate:2009_a,Endo:2010_a,Fang:2010_a,Miron:2010_a,Li:2013_a}, observed previously in broken inversion-symmetry, spin-orbit coupled paramagnets or FMs.   
We illustrate in Fig.~\ref{SOT_angle}(a) how 
these current induced non-equilibrium fields arise in structures with broken inversion symmetry. Here we choose the case of a Rasbha spin-orbit coupled 2D system for simplicity. The electric field induces an asymmetric non-equilibrium  distribution function of carrier eigenstates and as a result a net polarization ensues that depends on the scattering time, hence its link to extrinsic scattering origin. In magnets, the non-equilibrium carrier spin density acts on  magnetic moments as an effective magnetic field when carrier spins are exchange-coupled to the magnetic moments.

The full lattice of the Mn$_2$Au crystal has an inversion symmetry and the first expectation would be that there is no current-induced spin-density.
However, the lattice is formed by two sublattices which, individually, have broken inversion symmetry and form inversion partners  along the [001] axis. These coincide with the spin sublattices of the AFM ground-state in Mn$_2$Au, as highlighted in Fig.~\ref{Mn2Au}(a). 
The two sublattices forming the inversion partners in the Mn$_2$Au crystal are at the origin of the observed intra-band NSOT. 
 

{\it Results in the model 2D Rasba AFM:} Since both spin sublattices experience the same inversion symmetry breaking Rashba field in our 2D AFM model, the intra-band contribution to the current induced spin polarization has the same sign on both spin sublattices, i.e., is not staggered. A NSOT field is found, however, when evaluating the inter-band term $\delta \vec{s}^{inter}$ from Eq.~(\ref{inter}). 
The N\'eel-order current-induced field components projected on each sublattice are  shown in Fig.~\ref{SOT_angle}(e),(f)  for the AFM spin-axis rotation in the  [100]-[010] plane   and in the [100]-[001] plane. Here 
we plot the corresponding NSOT field per applied current along the [100]-direction.

The inter-band contribution described by Eq.~(\ref{inter}) arises from the time-dependent quantum-mechanical perturbation of the eigenstates between collisions (illustrated in  Fig.~\ref{SOT_angle}(d)) and is the basis of the  Berry curvature mechanism introduced to explain the intrinsic anomalous Hall effect \cite{Nagaosa:2010_a}, the intrinsic spin-Hall effect \cite{Jungwirth:2012_a}, and most recently also the intrinsic anti-damping spin-orbit torque in FMs \cite{Kurebayashi:2013_a,Freimuth:2013_a}. The key phenomenology which distinguishes the inter-band anti-damping spin-orbit field from the intra-band field is the harmonic dependence on the in-plane and out-of-plane spin-axis angles, as shown in  Figs.~\ref{SOT_angle}(e),(f). 
\begin{figure}[h]
\hspace*{0cm}\epsfig{width=.9\columnwidth,angle=0,file=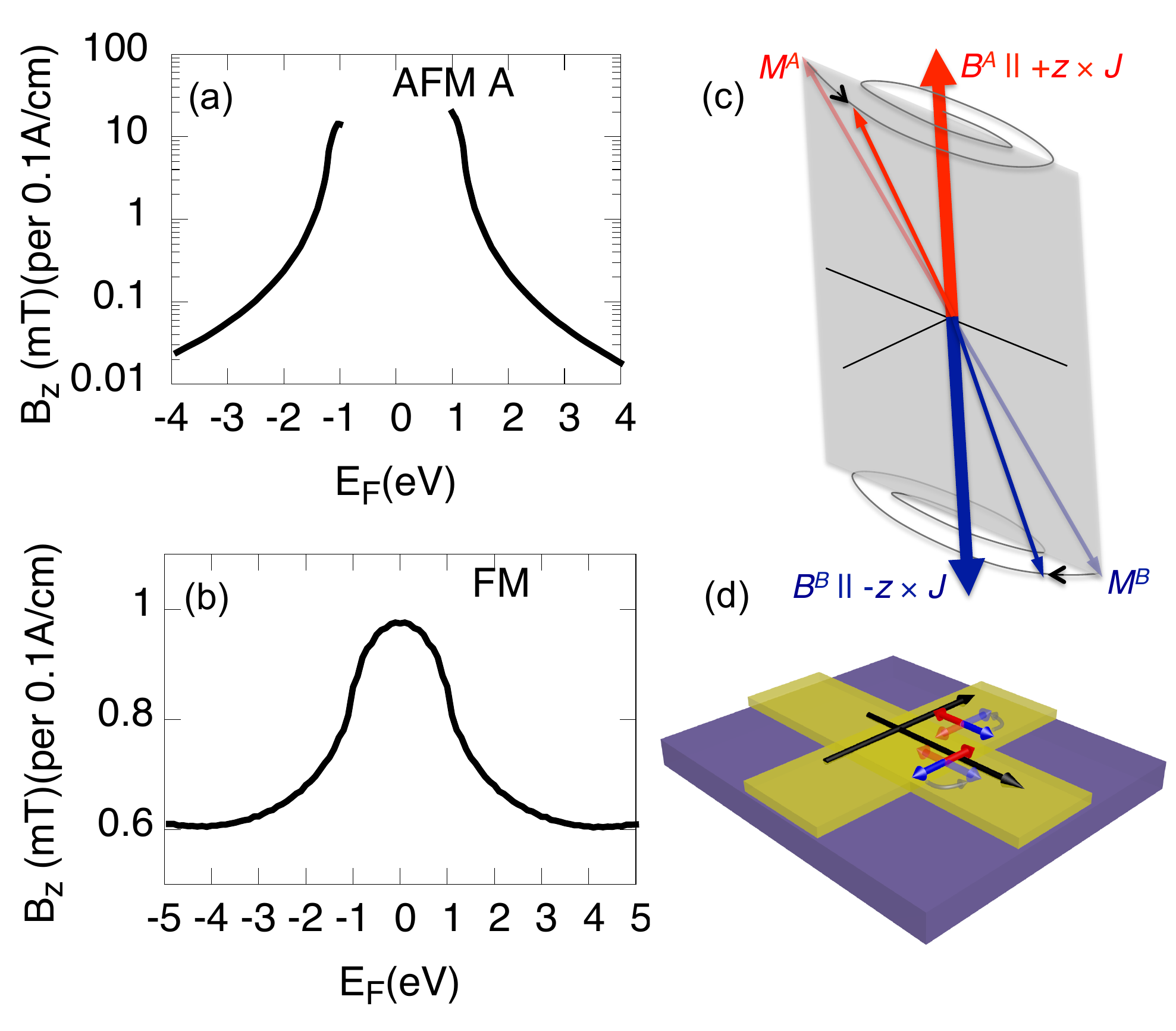}
%
\caption{(a) z-component of the NSOT field in the 2D square-lattice Rashba AFM for $\phi=0$ as a function of the Fermi energy. (b) Same as (a) for the FM state of the 2D Rashba square-lattice. (c) Schematics of the AFM dynamics in a staggered field generating the field-like torque. (d) Schematics of the electrical writing scheme via the NSOT in a memory device built in an AFM with cubic easy-axes. Black arrows represent the cross-wire writing currents and either of the current lines can be also used for reading by the AMR. Red-blue arrows show the current-induced AFM spin-axis reorientation.}
\label{SOT_EF}
\end{figure}
Another important feature, illustrated in Figs.~\ref{SOT_EF}(a),(b), is that the inter-band NSOT in the AFM can be significantly larger than its  FM SOT counterpart. The inter-band nature of the term $\delta \vec{s}^{inter}$ from Eq.~(\ref{inter}) implies that its magnitude is large when two sub-bands linked by spin-orbit coupling have a small energy spacing. In the calculations shown in Figs.~\ref{SOT_angle}(e),(f) and \ref{SOT_EF}(a),(b) for the model 2D lattice, we consider  weak spin-orbit coupling relative to the  exchange energy strength \cite{Kurebayashi:2013_a}. The smallest band splitting in this limit is governed by the exchange energy in the FM state while in the AFM state it is given by the small spin-orbit splitting. The band structures with their corresponding splittings are plotted in Fig.~\ref{2DRashba} together with the corresponding densities of states (DOSs) showing the characteristic van Hove singularities of the 2D square lattice.  As shown in Figs.~\ref{SOT_EF}(a),(b), the inter-band current induced spin-orbit fields are enhanced in the vicinity of the DOS singularities in both the FM and AFM states. However, in the AFM state with the small energy spacing of the two spin-orbit coupled bands, the enhancement is much larger, reaching three  orders of magnitude in the present calculations. 

{\em Discussion:} In our 2D Rashba model we identified a relativistic microscopic mechanism by which an electrical current ${\bf J}$ driven in a plane of an AFM layer  with broken space-inversion symmetry generates an anti-damping NSOT. It acts on the sublattice magnetizations ${\bf M}^A=-{\bf M}^B$, in the form  ${\bf T}^{A/B}\sim {\bf M}^{A/B}\times {\bf B}^{A/B}$, where ${\bf B}^{A/B}\sim[{\bf M}^{A/B}\times(\hat{z}\times{\bf J})]$. The effective field ${\bf B}^{A/B}$ is staggered in this case due to the opposite magnetizations on the two sublattices.  This is reminiscent of an earlier phenomenological prediction of a non-relativistic anti-damping spin transfer torque (STT) generated in an AFM by an effective field ${\bf B}^{A/B}\sim{\bf M}^{A/B}\times{\bf p}_{FM}$  due to a vertical spin-current  from an adjacent FM layer polarized along a vector ${\bf p}_{FM}$ \cite{Gomonay:2010_a}. A detailed study of the corresponding Landau-Lifshitz-Gilbert (LLG) dynamics (assuming a fixed ${\bf p}_{FM}$) showed that above a critical current, at which the energy loss due to internal damping is compensated by the current-induced pumping,  the configuration of ${\bf M}^{A/B}\parallel{\bf p}_{FM}$ becomes unstable and is switched to a stable ${\bf M}^{A/B}\perp{\bf p}_{FM}$ state \cite{Gomonay:2010_a}. Because this is independent of the sign of the vertical current,  the STT cannot switch the AFM back  to the ${\bf M}^{A/B}\parallel{\bf p}_{FM}$  configuration. In the case of our anti-damping NSOT,  ${\bf p}_{FM}$ is replaced with $\hat{z}\times{\bf J}$ and a reversible 90$^\circ$ reorientation of the AFM can be achieved by redirecting the in-plane current ${\bf J}$ between two orthogonal directions. Note that this favorable property of an anti-damping torque induced in the AFM by an in-plane current would apply not only to our relativistic NSOT but also to a spin-injection into the AFM from an adjacent paramagnet layer via the relativistic  spin Hall effect \cite{Miron:2011_b,Liu:2012_a}. 

The field-like STT acting on the AFM in the FM/AFM bilayer has the form ${\bf T}^{A/B}\sim {\bf M}^{A/B}\times {\bf B}^{A/B}$ with ${\bf B}^{A/B}\sim{\bf p}_{FM}$, i.e., ${\bf B}^{A/B}$ does not have the desired staggered property \cite{Gomonay:2014_a}. This illustrates why field-like non-relativistic STTs have been neglected in the LLG dynamics induced by an electrical current in AFMs. In our microscopic study of the Mn$_2$Au we have demonstrated, however, that a field-like torque  ${\bf T}^{A/B}\sim {\bf M}^{A/B}\times {\bf B}^{A/B}$ with a staggered ${\bf B}^A\sim +\hat{z}\times{\bf J}$ and ${\bf B}^B\sim -\hat{z}\times{\bf J}$ can be generated by current in special crystal structures in which the AFM spin sublattices coincide with the two inversion-partner sublattices. The corresponding dynamics comprising (damped) elliptical precessional motion  with opposite helicities of the two spin sublattices is sketched in Fig.~\ref{SOT_EF}(c). For a detailed description of these modes we refer to Eq.~(9) and the corresponding discussion in Ref. \cite{Keffer:1952_a}. Again, as in the anti-damping NSOT case, a cross-wire geometry can be used to reversibly switch between two orthogonal AFM spin-axis directions using the field-like NSOT.  For the theoretical (001)-plane anisotropy energies in Mn$_2$Au \cite{Shick:2010_a}, our estimates of the NSOT fields suggest sizable reorientations at current densities $\sim 10^8-10^9$~A/cm$^2$, depending on the angle of the applied in-plane current with respect to the easy and hard anisotropy axes.



Ultra-fast (ps-scale) reorientation of the AFM spin-axis in a staggered field was previously reported in an optical pump-and-probe study of AFM TmFeO$_3$ \cite{Kimel:2004_a}. The origin of the staggered field was different than considered here; the material had a  temperature dependent  AFM easy-axis direction and the corresponding N\'eel-order anisotropy field was induced by laser-heating the sample above the easy-axis transition temperature. The microscopic origin of the staggered field is not crucial, however, for the time-scale of the spin-dynamics which in AFMs is typically 2-3 orders of magnitude shorter than in FMs \cite{Keffer:1952_a}. We can therefore infer from these experiments  that  the AFM spin-axis reorientation due to our current-induced NSOT will not be limited by the ultra-fast AFM spin dynamics itself but only by the circuitry time-scales for delivering the electrical pulses which can be of order  $\sim100$~ps \cite{Schumacher:2003_a}. A schematic of a device that can be used to reverse the AFM spin-axis  electrically between two orthogonal directions is illustrated in Fig.~\ref{SOT_EF}(d). In a cross-wire geometry, each wire stabilizes via the NSOT one of the two orthogonal spin-axis directions. An AFM-AMR \cite{Marti:2014_a,Marti:2013_a} measured in one of the arms can then be used to electrically detect the spin-axis direction. In an AFM with cubic magnetic anisotropy, an all-electrical AFM memory device can be realized based on this scheme.

We acknowledge  fruitful discussions with R. Campion, K. Edmonds, A. Ferguson, B. Gallagher, X. Marti, V. Novak, K. Olejnik, H. Reichlova, A. Rushforth, H. Saidaoui, and P. Wadley, and  support from the EU European Research Council (ERC) advanced grant no. 268066, from the Ministry of Education of the Czech Republic grant no. LM2011026, from the Grant Agency of the Czech Republic grant no. 14-37427G,  from the Academy of
Sciences of the Czech Republic Praemium Academiae, from the NSF grant no. DMR-1105512, ONR grant no. 141110780, and the Alexander Von Humboldt
Foundation. A.M. was supported by the King Abdullah University of Science and Technology

\smallskip
$^\ast$ The authors contributed equally.


\end{document}